\documentstyle[twoside,fleqn,espcrc2]{article}

\input BoxedEPS
\SetRokickiEPSFSpecial  %% for dvips by Tom Rokicki, for VMS
\HideDisplacementBoxes

\newcommand{\staple} {\mbox{
\hspace{-0cm}\raisebox{.22cm}{$\scriptscriptstyle\leftarrow$} 
\hspace{-.48cm}\raisebox{.22cm}{$\scriptscriptstyle-$}   
\hspace{-.30cm}\raisebox{-.08cm}{$\scriptscriptstyle\rightarrow$}
\hspace{-.185cm}\raisebox{-.08cm}{$\scriptscriptstyle-$} 
\hspace{-.215cm}\raisebox{-.00 cm}{$\scriptscriptstyle\uparrow$} 
\hspace{-.222cm}\raisebox{.13 cm}{$\scriptscriptstyle |$}  
}}
\newcommand{\link} {\mbox{
\hspace{-0cm}\raisebox{-.00 cm}{$\scriptscriptstyle\uparrow$} 
\hspace{-.222cm}\raisebox{.13 cm}{$\scriptscriptstyle |$}  
}}
\newcommand{\boxmn} %%% Fiddled with the numbers -- Marty
{\setlength {\fboxrule}{.1mm} \setlength {\fboxsep}{.2mm}%
 \fbox{%
  \rule[0mm]{0mm}{2.4mm} 				      %control height
  \rule {1mm}{0mm}               %space
  \raisebox{-.35mm}{$\scriptscriptstyle m$} 
  \rule {-1.2mm}{0mm}               %space
  \raisebox{.6mm}{$\scriptscriptstyle n$}
  \rule {-2.5mm}{0mm}               %space
 }
} 

\newcommand{\Tr}{\mathop{\rm Tr}}  %%%  This is how to make up
\renewcommand{\Im}{\mathop{\rm Im}}  %%% new operators.
\newcommand{\Det}{\mathop{\rm Det}}
\newcommand{\AmS}{{\protect\the\textfont2
  A\kern-.1667em\lower.5ex\hbox{M}\kern-.125emS}}

\title{Instantons, the QCD Vacuum, and Hadronic Physics
\thanks{This work is supported in part by the U. S. Department of Energy
(DOE) under cooperative research agreement \#DE-FC02-94ER40818.}}
\author{J. W. Negele 
\address{Center for Theoretical Physics, 
Laboratory for Nuclear Science, and Department of Physics, \\
Massachusetts Institute of Technology, 77 Massachusetts Ave., Cambridge,
MA 02139, USA}}
\begin{document} 

\begin{abstract}

 A large body of evidence from lattice calculations
indicates that instantons play a major role in the physics of light
hadrons.  This evidence is summarized, and recent results concerning the
instanton content of the SU(3) vacuum, instanton contributions to the
static potential, and a new class of instanton solutions at finite
temperature are reviewed. 

\end{abstract}

\maketitle

\section{INTRODUCTION AND PHYSICAL PICTURE}

\subsection{Physics motivation}

One of the great advantages of the path integral formulation of quantum
mechanics and field theory is the possibility of identifying
non-perturbatively the stationary configurations that dominate the action
and thereby identify and understand the essential physics of complex
systems with many degrees of freedom.  Thus, the discovery of instantons
%nearly a quarter of a century ago
in 1975~\cite{belavin+75} gave rise to great
excitement and optimism that they were the key to understanding QCD. 
Indeed, in contrast to other many body systems in which the quanta
exchanged between interacting fermions can be subsumed into a potential, it
appeared that QCD was fundamentally different, with topological
excitations of the gluon field dominating the physics and being
responsible for a host of novel and important effects including the
$\theta$ vacuum, the axial anomaly, fermion zero modes, the mass of the
$\eta^\prime$, and the chiral condensate.  However, despite nearly a
quarter of a century of theoretical effort, it has
not been possible to proceed analytically beyond the dilute instanton gas
approximation~\cite{callan+78a}. In the intervening years, the  
instanton liquid  model~\cite{shuryak82,diakonov+86,schafer+98} provided a
successful phenomenology and qualitative physical understanding, but a 
quantitative exploration of the role of instantons in nonperturbative QCD
has had to wait until lattice QCD became sufficiently sophisticated and
sufficient resources could be devoted to the study of instanton physics. 

In recent years, significant progress has been made by a number of groups
in understanding instanton physics on the lattice, and this review
summarizes some of the highlights. This field is both exciting and
frustrating. It is exciting because, at last,  a substantial body of
evidence shows that the original vision of the role of instantons in QCD is borne
out in nature. However, as will be clear below, it is also frustrating because of
intrinsic limits to the precision with which one can apply ultimately semiclassical
concepts. Since there is more interesting material than could be accommodated, I
will omit significant developments in spectral methods, which are covered in
another talk~\cite{narayanan98}, and recent work on abelian projection and
monopoles, a subject to which I am not able to provide any additional insight. 

\subsection{Aspects of continuum instanton \hfill\\physics}

To put lattice investigations in context, it is useful to recall relevant
aspects of continuum instanton physics. Working in Euclidean time, we
evaluate a path integral of the form $\int{\cal D}[A]e^{\int d^4x\, S[A]}$.
Hence, as in statistical mechanics, the weight of a configuration depends
not only on its energy, but also on its entropy -- %%% --, not -, here
 the number of ways it
can be realized. In addition, tunneling solutions arise as periodic
classical solutions in an inverted potential. Vortices in two-dimensional
spin systems provide a useful analogy.  Classified by
winding number, the cost of the localized core energy is compensated by
entropy so that vortices can become the dominate degrees of freedom as in
the Kosterlitz-Thouless phase transition. Similarly, we may expect BPST
instantons~\cite{belavin+75} which connect degenerate minima of differing
winding number with the self-dual gauge potential
$A_{\mu}^a(x) = {{2 \eta_{a\mu\nu}x_{\nu}}\over{x^2 + \rho^2}}$ having
scale-invariant action $S = {1\over 4}\int d^4x\, F_{\mu\nu}^a
F_{\mu\nu}^a = {8\pi^2 g^{-2}} \equiv S_0$ to have a significant
presence in the vacuum due to the high entropy associated with translation,
size, and color orientation. The action and topological charge density are
localized around the center as $\pm F_{\mu\nu}^a \tilde F_{\mu\nu}^a =
F_{\mu\nu}^aF_{\mu\nu}^a = {192\rho^4\over (x^2+\rho^2)^4}$\ . The
%F_{\mu\nu}^aF_{\mu\nu}^a = {192\rho^4 (x^2+\rho^2)^{-4}}$\ . The
tunneling rate is $dn_I \sim ({8\pi^2\over
g^2})^{\scriptscriptstyle 2N_c}{d\rho\over
\rho^5} d^4x^{(I)} e^{-{8\pi^2\over g^2(\Lambda^{-1})}} (\Lambda \rho)^{11
N_c\over 3} $, where the prefactor
and the running of the coupling constant in the last factor  produce a distribution
of instantons $\sim \rho^6$ for SU(3).  Physically, we expect this distribution to be
cut off at large
$\rho$ by interactions between instantons and by fluctuations when the amplitude
of a sufficiently large instanton becomes small relative to  quantum fluctuations.
It is the difficulty in treating these infrared effects that has stymied analytic
progress.

From the axial anomaly, \\
$\partial_{\mu}\sum_f\bar\psi\gamma_{\mu}\gamma_{5}\psi =$
$2m\sum_f\bar\psi
\gamma_5\psi + {N_f\over 16 \pi^2}F_{\mu\nu}^a \tilde F_{\mu\nu}^a$,\\
the topological charge satisfies the index theorem and, for periodic
systems, may be expressed in terms of fermion eigenfunctions \\
$Q\!=\!{g^2\over
32\pi^2}\int F\tilde F% F_{\mu\nu}^a \tilde F_{\mu\nu}^a
=n_L-n_R=m\sum_{\lambda}{{\int \psi_{\lambda}^{\dag}(x)
\gamma_5\psi_{\lambda}(x)\over m+i\lambda}} $,
 where $n_L$ and $n_R$ denote the number of fermion zero
modes.  For an isolated instanton, the zero mode is 
$\psi_0(x)={{\rho\gamma\cdot\hat
x(1+\gamma_5)}\over2\pi(x^2+\rho^2)^{3/2}}\phi$. In the limit of light quarks,
the Greens function for N$_f$ quarks reduces to the product of zero modes \\
$\prod_f \Det[ {D\!\!\!\!/}+m] \bar\psi_f(x)\psi_f(y)
{\;\longrightarrow _{\!\!\!\!\!\!\!\!\!\!\!\!\!\scriptscriptstyle {
\atop m\rightarrow 0}}}
\prod_f \psi_0^{ }(x) \psi_0^{\dag}(y)$\\ and gives rise to the 't Hooft
interaction. Thus, light quarks propagate by zero modes which
in turn arise from instantons.  Based on large N arguments, the
Veneziano-Witten formula~\cite{witten79,veneziano79} relates the
$\eta^{\prime}$ mass to the topological susceptibility in the pure gluon
sector\\
$\chi \equiv \int {d^4x\over V} \langle Q(x)Q(0)\rangle  = %% \langle not <
{f_{\pi}^2\over2N_f}(m_{\eta}^2 + m_{\eta^{\prime}}^2 -2m_K^2)$\\
yielding the expectation that $\chi = (180\, MeV)^4$. 

Finally, for an
ensemble of instantons and anti-instantons, we expect mixing of their zero
modes to generate a finite density of states near zero virtuality.  Hence,
by the Banks-Casher relation~\cite{banks+80}, \\
$-\langle\bar\psi\psi\rangle = \int d\lambda\,\rho(\lambda) {2m\over
\lambda^2+m^2} {\;\longrightarrow _{\!\!\!\!\!\!\!\!\!\!\!\!\!\scriptscriptstyle {
\atop m\rightarrow 0}}} \pi\rho(\lambda\!=\!0)$, they account for the chiral
condensate.  Subsequent sections will
discuss the extent to which this continuum instanton physics can now be
seen on the lattice.

\subsection{The instanton liquid}

The instanton liquid model~\cite{shuryak82,diakonov+86,schafer+98} provides
an economical phenomenology of instanton mediated quark propagation in the
QCD vacuum. The integral over all gluon fields that one evaluates in lattice 
QCD using an ensemble of configurations sampling the action is replaced by
an  ensemble of instanton and anti-instanton configurations.  Although the
model has been refined to account for interactions between instantons and
the fermion determinant, the gross features can be seen for the case of an
ensemble of instantons and anti-instantons %%% was <antiinstonton
 of size $\rho \sim 1/3$\,fm and
density $n\sim 1\,\mbox{fm}^4$ randomly distributed in space
and color orientation,  where the values of $\rho$ and $n$ are determined
from the physical gluon and chiral condensates. 

Due to the opposite shifts
of left and right-handed fermion levels in the presence of an instanton,
one may think of the 't Hooft interaction as a vertex that absorbs
left-handed particles of each flavor and creates corresponding
right-handed particles, and {\it vice versa} for anti-instantons. Mesons
then propagate in the QCD vacuum by the hopping of quark-antiquark
pairs between these vertices, and the qualitative features of the
channel dependence arises naturally. Considering two flavors
for simplicity, a spin-zero pion propagates by a $u_R\bar d_L $ pair
interacting with an anti-instanton to produce a $u_L\bar d_R$ pair which in
turn interacts with an instanton to return to a  $u_R\bar d_L $ pair. 
Since the 't Hooft vertex is most attractive in this channel and the
interaction can act in all orders, the pion is the most strongly
attractive meson channel. In contrast, the spin-one rho propagates only in
second order, with an anti-instanton taking a $u_R\bar d_R $ pair to a 
$u_L d_L \bar d_R\bar d_R $ state which is then returned to a $u_R\bar d_R $
pair by an instanton, so we expect the interaction in this channel to be
much weaker.  Similarly, the scalar meson channel is repulsive and one
expects the nucleon channel to be more attractive than the delta channel.
The chiral condensate arises naturally in this picture by the fact
that the zero modes for isolated instantons mix in the instanton liquid
giving rise to a finite density of states at low virtuality.   

The tendency
of instanton--anti-instanton pairs connected by light quarks and antiquarks
to form dipoles oriented in the thermal direction suggests a possible
mechanism for chiral symmetry restoration at high temperature and may be
manifested by dipoles wrapped around the periodic thermal direction above
$\mathrm{T_C}$.

\section{EVIDENCE FOR INSTANTONS IN HADRON STRUCTURE}

The expectations raised by the our analytical understanding of the role of
instantons in QCD and the physical picture of quarks hopping between zero
modes associated with instantons have now been borne out by a broad range
of lattice calculations as described below. 

\subsection{Vacuum current correlation functions}

Vacuum correlation functions for space-like separated hadron
currents calculated in lattice QCD display the qualitative
behavior expected from the 't Hooft interaction and agree
semi-quantitatively with the instanton liquid model. As emphasized in
ref~\cite{shuryak93}, correlation functions of the form $R(x) \equiv \langle 0|T
J_{\mu}(x) J_{\mu}(0)|0\rangle $ characterize the spatial and channel dependence
of the interaction between quarks and antiquarks and thus supplement hadron
bound state properties like phase shifts supplement deuteron properties in
characterizing the nuclear interaction. The ratio
$R(x)/R_0(x)$ of the interacting to free correlator has been calculated in
quenched QCD for the following meson and baryon currents~\cite{chu+93}, $J = 
\bar u\gamma_{\mu}d,
\bar u \gamma_{\mu}\gamma_5 d, \bar u\gamma_5 d, \bar u d,
\epsilon_{abc}[c^aC\gamma_{\mu}u^b]\gamma_{\mu}\gamma_5d^c,$
and $\epsilon_{abc}[u^a C\gamma_{\mu}u^b]u^c$.  
Results are consistent with dispersion analysis of e$^+$-e$^-$
and other data in relevant channels, and typical results are shown in Fig.~1. 

\begin{figure}[t]
\vspace*{-.2cm}
$$
\BoxedEPSF{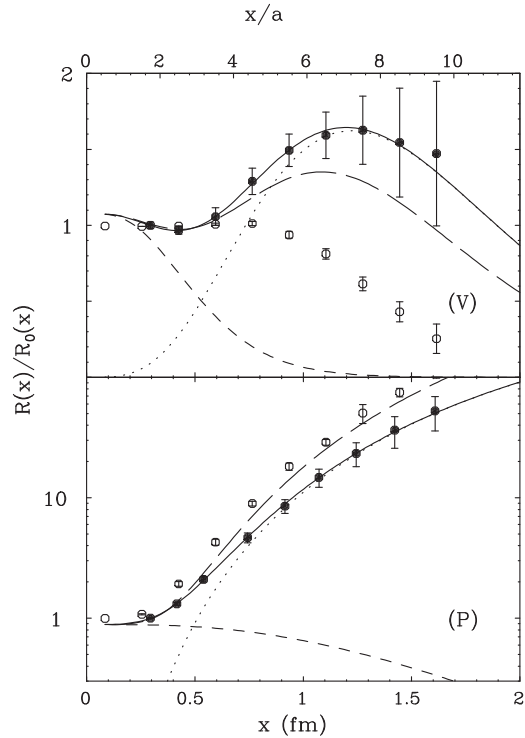 scaled 850}
$$
\vspace*{-1cm}
\caption{Vector $(V)$ and 
Pseudoscalar $(P)$ correlation functions from lattice calculations
\protect\cite{chu+93} (solid circles), from the random instanton
model \protect\cite{shuryak+93,schafer+94} (open circles) and from dispersion
analysis
\protect\cite{shuryak93} (long dashes). The solid curve fit to the lattice
data is made up of a continuum contribution (short dashes) plus a resonance
contribution (dotted curve).}
\label{f1} 
\vspace*{-.6cm} 
\end{figure}

Note for subsequent reference, that the solid curve fit to the lattice
data may be decomposed into a continuum contribution concentrated near the
origin (short dashed line) and a resonance contribution arising from the
rho or pion which dominates in the region of 1 fm and beyond (dotted line). 
Quenched calculations at $\beta = 6.2$~\cite{hands+95} corroborate the
original
$\beta = 5.7$ results. 

\subsection{Comparison of results with all gluons and only instantons}

\begin{figure}[t]
\vspace*{-.2cm}
$$
\BoxedEPSF{%new-axial-xmgr-bin_bw.eps scaled 400
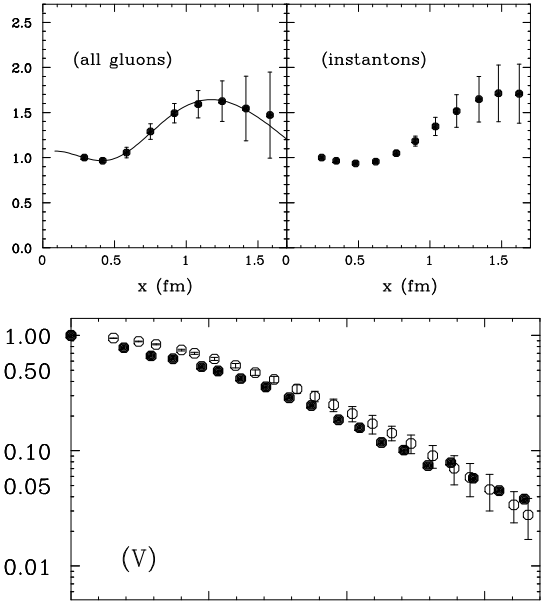 scaled 666} 
$$
\vspace*{-1.2cm}
\caption{Vacuum
correlator in the rho channel calculated with all gluons (upper left) and
with only instantons (upper right)~\cite{chu+94}.  The rho meson ground state
density-density correlation function calculated with all gluons (solid
circles) and with only instantons (open circles) is shown below.}
\label{f2}  
\vspace*{-.5cm}
\end{figure}

One dramatic indication of the role of instantons in light hadrons is to
compare observables calculated using all gluon contributions with those
obtained using only  the instantons remaining after cooling. The basic idea
of cooling~\cite{berg81,teper91}, and the variants described in Section 3, is to 
iteratively locally relax a lattice gluon configuration and thereby approach the
nearest local minimum of the classical action.  Thus, one may extract the
instanton content by removing local fluctuations while preserving
topologically %%% was <toplologically>
 stable excitations. When a
$\beta = 5.7$ quenched  configuration is cooled 25 steps, the gluon content
changes dramatically~\cite{chu+94}: the fluctuations in the action and
topological charge densities decrease two orders of magnitude,  self-dual
instantons and anti-instantons become apparent,
$\langle S\rangle\over S_0$ decreases from over 20,00 to 64, the string tension
reduces to one fourth its original value, the lattice spacing defined by
the nucleon mass decreases from 0.168fm to 0.142fm, and the magnetic
hyperfine components of the quark-quark potential become essentially zero. 
Hence, for example, the energies and wave functions of charmed and $B$
mesons would be drastically changed.

As shown in Fig.~\ref{f2}, however, the properties of the rho meson are
virtually unchanged.  The vacuum correlation function in the rho (vector)
channel and the spatial distribution of the quarks in the rho ground state,
given by the ground state density-density correlation
function~\cite{chu+91} $\langle \rho| \bar{q} \gamma_0 q(x) \bar{q}
\gamma_0 q(0) | \rho \rangle$, are statistically indistinguishable before
and after cooling.  Also, as shown in Ref.~\cite{chu+94}, the rho mass is
unchanged to within its 10\% statistical error.  In addition, the pseudoscalar,
nucleon, and delta vacuum correlation functions and nucleon and pion
density-density correlation functions are also qualitatively unchanged after
cooling, except for the removal of the small Coulomb induced cusp at the
origin of the pion. Similarly, the
axial charge matrix elements specifying the spin content of the nucleon,
$\langle \vec{P}\vec{S}|(\bar{q} \gamma^{i} i \gamma_5
q)|\vec{P}\vec{S}\rangle = 2 S^i\Delta q$, are quite similar when calculated
with all gluons and only instantons~\cite{dolgov+98}. This result suggests
that when all the quark and gluon contributions are calculated,
instantons, which are the natural mechanism to remove helicity from the
valence quarks and transfer it to gluons and sea quarks, may account for the
proton ``spin crisis''.

\begin{figure} [tp]%         [tp]
$$
\BoxedEPSF{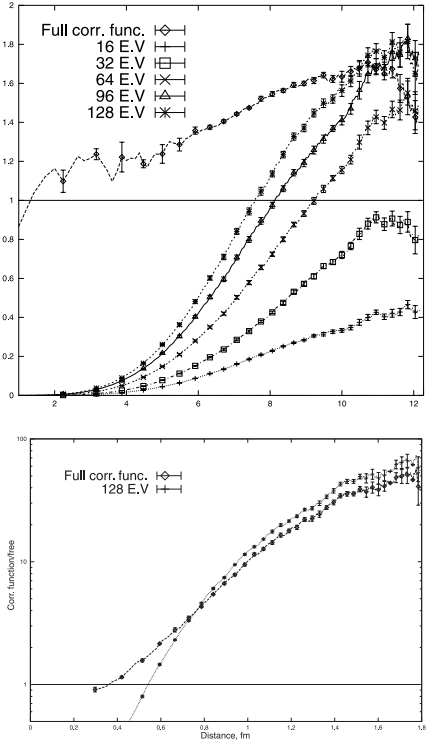 scaled 621}  
%\BoxedEPSF{Lat98Fig3.eps scaled 777}  
$$
\vspace*{-13mm}
\caption {Truncated eigenmode expansion~\cite{ivanenko+98}.}
\label{f3}  
\vspace*{-5.mm}
\end{figure}

\subsection{Zero mode zone expansion}
 
Because, as discussed in Section 3, cooling somewhat modifies the
instanton distribution, it is useful to study the zero modes of the original
uncooled configuration directly. Several
studies~\cite{narayanan98,ivanenko+98,venkataraman+97,gattringer+98} calculate
zero modes on the real axis and conjugate pairs displaced slightly off the axis due
to the mixing of instantons and antiinstontons, examine the discrete lattice
counterpart of the index theorem, and consider the presently unsolved problem of
separating physical continuum solutions from doublers. A clear indication of the
role of zero modes in light hadron observables is the degree to which truncation of
the expansion $ \sum_{\lambda} {\psi_{\lambda}(x)
\psi_{\lambda}(y)^{\dagger}\over m+i \lambda}$ to the zero mode zone
reproduces the result using the complete propagator. The top plot in
Fig.~\ref{f3} shows the effect of including the lowest 16
through 128 of the 786,432 modes  on a 16$^4$ lattice in the vector
correlation function, indicating that already 64 modes reproduce most of the
strength of the rho peak and 128 modes produce the full strength.  The
lower plot shows that 128 modes also reproduce the full pion
strength~\cite{ivanenko+98}. Similarly, most of the strength of the disconnected
graph contribution to the $\eta^{\prime}$ correlation function, which should be
particularly sensitive to instantons, is already provided by the lowest 32
eigenmodes, and the fermionic definition of the topological charge is nearly
saturated by the lowest 8 modes~\cite{venkataraman+97}.

\subsection{Localization of zero modes}

\begin{figure}[tp]
\vspace*{-3mm}
$$
\BoxedEPSF{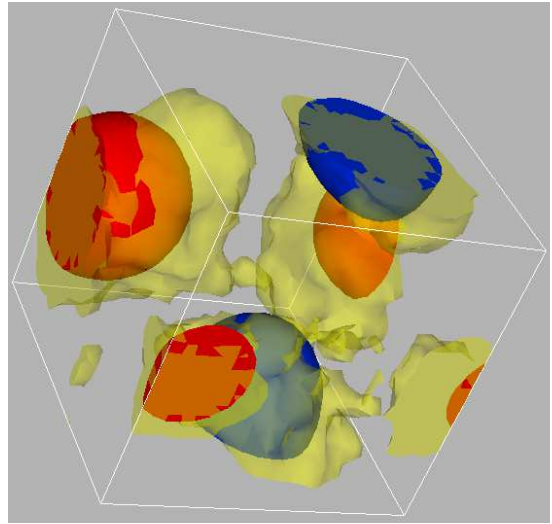 scaled 366} 
$$
\vspace*{-3pc}
\caption{ Isosurfaces on a periodic lattice of topological charge density
$F\tilde F(x)$ calculated from a cooled configuration (dark surfaces) and
$\bar\psi_0(x)\psi_0(x)$ calculated on the corresponding uncooled
configuration (light surfaces)~\cite{hetrick98}.}
\label{f4}  
\vspace*{-.4cm} 
\end{figure}

The observation of localization of low eigenmodes of {\it uncooled}
configurations at the locations of instantons identified by cooling confirms
several important ideas. Even in the presence of fluctuations several orders
of magnitude larger than the instanton fields themselves, the light quarks
essentially average out these fluctuations and produce localized peaks at
the topological excitations as expected from semiclassical arguments. In
addition, it is clear that the instantons which survive cooling
represent 
topological %%% was <topoligical>
structure in the original configuration. This
localization has been observed for unquenched Wilson fermions at T = 0,
$\beta$ = 5.5~\cite{ivanenko+98} and for unquenched staggered fermions at t
$\le$ T$_c$ at $\beta=5.65$~\cite{hetrick98}.  Isosurfaces of
$\bar\psi_0(x)\psi_0(x)$ for the lowest uncooled
eigenmode~\cite{hetrick98} are compared with the cooled topological charge
density in Fig.~\ref{f4}.

\subsection{Molecular correlations}
 
At high temperature, the fermion determinant favors alignment of
instanton--anti-instanton pairs in the thermal direction~\cite{shuryak93}.
Although aligned molecules have not been observed directly in the
topological charge density of cooled configurations, a suggestive anisotropy
in the topological susceptibility measured on differently oriented
sublattices was reported at this conference~\cite{hetrick98}. Lattice slices
that are narrow in the time direction but span the full spatial volume, 
which would have a large range in the value of Q for a dipole layer, were
observed to have
$\langle Q^2\rangle$ up to twice as large as slices that are narrow in one spatial
direction but span the full time range, which would tend to have canceling
positive and negative charges from the dipoles in the slice.  Whereas it is
not presently understood why a comparable asymmetry is also observed 
in quenched SU(2), subvolume susceptibility measurements are an interesting
and promising development.

\subsection{Spectral density
 and \protect\boldmath$\langle\bar\psi \psi\rangle$}%%% bold math in heading

Calculations of the lowest 8 eigenvalues for a set of MILC configurations
have been used to evaluate the density of states at low virtuality,
$\rho(0)$, and thereby numerically check the Banks-Casher
relation on the lattice~\cite{hetrick98}. The results of low $\lambda$ fits
to the integrated histograms yield values of 
$\langle\bar\psi\psi %%% \langle, not < , etc. 
\rangle_{m_q\rightarrow 0}$ 
(to be compared with
${\pi\over V} \rho(0)$ in parentheses) of 0.0114 (0.0138) at $\beta$ = 5.65 and
0.002 (0.002) at $\beta = $ 5.725. 

The significant developments in the use of domain wall chiral fermions have
resulted in a number of studies of zero modes reported at this
conference~\cite{blum98}.  For sufficiently large separation between domain walls,
there is an integer index for the lattice Dirac operator corresponding to the
winding number for smooth fields as well as exact zero
modes~\cite{narayanan+95}. In the presence of zero modes, the chiral condensate
becomes $-\langle\bar q q\rangle={1\over V}\sum_{\lambda=0}{1
\over m} + {1\over V}\sum_{\lambda > 0} {2m\over\lambda^2+m^2}$, leading to
a $1\over m$ mass dependence. Recently~\cite{pchen+98}, calculations  with
domain walls separated by 10 on a 16$^4$ lattice in the presence of a smooth
configuration with winding number one plus random fluctuations demonstrated
this $1\over m$ dependence over the range $10^{-5}< m < 10^{-3}$. 
 
\section{INSTANTON CONTENT OF THE QCD VACUUM}
Several recent studies of the instanton content of the SU(3) vacuum provide
an unprecedented, albeit imperfect, view of the size distribution of
instantons and their spatial correlations. To appreciate the issues involved
in determining the instanton content, it is useful to briefly review the
methods that have been used. 
 
\subsection{Methods to extract the instanton\hfill\\ content}

Standard cooling~\cite{berg81,teper91} applied to QCD minimizes the action
locally at each link. For the Wilson action, one calculates 
 $\mathrm Max_{[U]}(Re \, Tr\, U \Sigma^{staples})$ so
that U$^{-1}$ is the projection on SU(N) of $\mathrm\Sigma^{staples} \!=
\sum_{\mu}$
\staple. In the case of SU(2), since the sum of staples is a multiple of a
group element, cooling makes the replacement \link $\rightarrow \sum_{\mu}$
\staple. For SU(3), it is convenient to sequentially minimize with respect
to SU(2) subgroups~\cite{cabibbo+82}. 

Underrelaxed cooling~\cite{michael+95} takes a linear combination of the
original link and the sum of staples, \link $\rightarrow (1-c) \link +
{c\over 6}\sum_{\mu}$ \staple, which corresponds to APE
smearing~\cite{falcioni+85} and for small $c$ has the effect of making the
minimization independent of the order of local updates. This has been used
in Ref.~\cite{smith+98} with $c$ =0.86 and sequential application to SU(2)
subgroups to analyze the UKQCD SU(3) configurations. This work also
calibrates the cooling at each mixing parameter $c$  and $\beta$  to
define a number of calibrated sweeps which produces approximately equivalent
evolution of instantons and instanton--anti-instanton pairs. Instantons are
identified by peaks in the topological charge density, Q(x), assuming
additivity and filters are applied to assure the appropriate shape and to avoid
counting fluctuations. 

Renormalization 
 group cycling, as originally implemented~\cite{degrand+97}, is
based on the renormalization group transformation $  e^{S(U)} =  \int d[V]
e^{S(V) + \kappa T(U,V)} $. The blocking transformation $\kappa T(U,V)$  maps
a configuration V on a fine lattice to a configuration U on a lattice twice
as course, and in the weak coupling limit the inverse transformation is given
by the saddle point condition Min$_{\{V\}}[S(V)+\kappa T(U,V)]$. Making use
of the fact that V is the smoothest configuration that blocks to U and that
the renormalization group preserves the instantons and large scale structure
of the configuration, the cycling procedure first inverse blocks from the
course lattice to the fine lattice and then reblocks back to a course
sublattice shifted one lattice spacing along the diagonal.

 Although cycling has many conceptual and
practical advantages, including stabilizing instantons 
with $\rho >0.94a$ and %% 0.94  not   .94
smoothing configurations sufficiently to identify instantons before the
string tension has been severely diminished, it is very expensive computationally.
Hence, efficient parameterizations of the cycling transformation
were studied~\cite{degrand+98} in SU(2).  After investigating 56 combinations
of paths up to 9 links, it was determined that a good parameterization was
obtained by APE smearing, providing an additional rationale for this
procedure. In addition, for reasons that are not yet clear, this work showed
that although the sizes of smooth instanton configurations are stable under
APE smearing, instantons and anti-instantons found in a configuration
generated by the Wilson action may change size (generally growing) linearly
with the number of steps. In  Ref.~\cite{ahasenfratz+98}, APE smearing
\link$\rightarrow [(1-c)\link + {c\over 6}\sum_{\mu} \staple ]_{SU(3)}$ with
direct projection onto SU(3) was used with c = 0.45 and 15-30 smearing steps
to study the vacuum distribution of instantons. For comparison,
UKQCD used c =  0.86 and 23-46 steps and the calculations of
topological charge by the Pisa group discussed below used c = 0.9 and 2
steps. 
 
Instead of sequentially minimizing the Wilson action, improved
cooling~\cite{deforcrand+97} minimizes an improved action constructed from
n$\times$m plaquettes,
$S = \sum_{\{n,m\}} c_{n,m} S_{n,m}$, where $S_{n,m} \equiv {1\over n^2m^2}
\sum_{x,\mu,\nu}\Tr (1 - \boxmn \,)$.  Up to five terms are included,
corresponding to 1$\times$1, 1$\times$2, 1$\times$3, 2$\times$2, and
3$\times$3 plaquettes, and the coefficients $c_{n,m}$ are determined to
eliminate leading corrections in $a$ and also to create a slight barrier to
prevent shrinkage of instantons above a critical size.  Whereas the
Wilson action for a smooth instanton monotonically decreases with $\rho$,
leading to the eventual shrinkage and disappearance of instantons after
hundreds of cooling steps, the five-term improved action is extremely flat
and preserves all instantons greater than $\rho_c$=2.3a indefinitely.
A comparably improved topological charge operator yields nearly integer
charge after 5-10 cooling steps. Improved cooling has also recently been
used to study the SU(3) vacuum~\cite{deforcrand+98}. 

A cooling method~\cite{ivanenko+98} that is particularly efficient for a
parallel computer is simultaneous relaxation of the link variables at all
lattice sites by discretizing the relaxation equation ${dU\over d\tau} =
-{\delta S\over \delta U}$.  
Parameterizing an SU(3) matrix in the vicinity of U with by $U[F]= e^{iF}U$,
where H is hermitian and traceless, and taking a small step $\Delta \tau$
yields\\ $U^{\prime} = e^{-i\Delta\tau[{1\over 2}UU_S-{1\over
2}(UU_S)^{\dag}-{1\over N_C}\Im \Tr(U U_S)]}U$,\\ where $U_S$ denotes the 
staple (or generalized staple for an improved action) associated with U.
Sufficiently small
$\Delta\tau$ produces  cooling histories comparable to overrelaxed
cooling, and  $\Delta\tau\!=\!0.025$ was used to analyze SU(3)
configurations.

\begin{figure}[tp]
\vspace*{-2mm}
$$
\BoxedEPSF{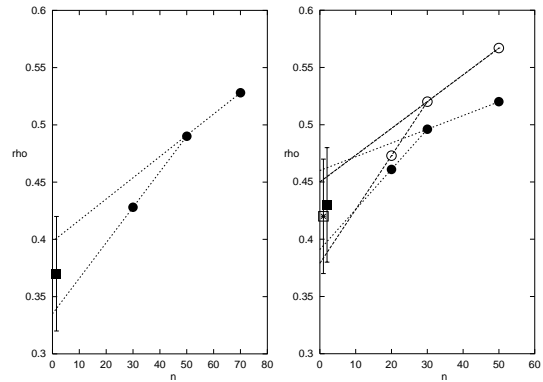 scaled 666}  
$$
\vspace*{-3pc}
\caption{Sketch of extrapolations of the average
instanton size to n$_c = 0$ cooling steps for data from 
Refs.\,\protect\cite{smith+98} (left) and 
\protect\cite{ivanenko+98} (right). Solid and open symbols denote quenched and
full QCD respectively. }
\label{f8}  
\vspace*{-.5cm} 
\end{figure}

\begin{table*}[tb]
\setlength{\tabcolsep}{1pc}
\newlength{\digitwidth} \settowidth{\digitwidth}{\rm 0}
\catcode`?=\active \def?{\kern\digitwidth}
\caption{Studies of the instanton content of the SU(3) Vacuum}
\label{t1}
\begin{tabular*}{\textwidth}{|c|@{\extracolsep{\fill}}c|l|c|c|c|l|}
\hline\
$\beta$ & Lattice & Method & $\bar\rho_{\rm N/V} $ &
$\bar\rho_{\rm extrap}$ & ${N\over V}^{\mathstrut}$& Ref.\\[0.5ex]
 & & &   (fm) &   (fm)  &(fm$^{-4}$) & \\[0.3ex]
\hline
 6.0? & $16^3\times48$ & Underrelaxed & 0.60(5)$^a$
& 0.37(5)$^b$ & 55--3.2 & \cite{smith+98}  ${}^{\mathstrut}$  \\
      & $32^3\times64$ & Cooling &  &  & &  \\
 6.2?  & $24^3\times48$  & & & & & \\
 6.4?  & $32^3\times64$  & & & & & \\
\hline
 5.85  & $12^4$ & APE Smearing &   & 0.32$^c$ &1.1& \cite{ahasenfratz+98} \\
 6.0?  & $12^4, 16^4$  & & & & &  \\
 6.1?  & $16^4$  & & & & &  \\
\hline
 5.85  & $12^4$ & Improved & $<0.53(5)^d$ &   & 3.3--0.38 &
\cite{deforcrand+98}
\\
 6.0?  & $16^4$ & Cooling &        &       &         & \\
\hline
 5.7?  & $16^3\times24$ & Cooling &  & $>0.39^e$ & 0.59--0.28$^f$    
&\cite{chu+94} 
\\
%[2ex]
 & & & & & & \\
\hline
 5.85 & $16^4$ & Relaxation & 0.50(5)$^g$ & 0.43(5)$^{b,g}$ & 5.3--1.4$^g$
&\cite{ivanenko+98}
\\
\cline{1-2}
\multicolumn{2}{|c|}{unquenched, $\kappa$=0.16} & & & & & \\
 5.5?  & $16^4$  &  & 0.52(5)$^g$ & 0.42(5)$^{b,g}$ & 6.5--1.8$^g$ &\\
\hline
\hline
\multicolumn{3}{|c|}{Summary} & 0.54(5) & 0.39(5) &\multicolumn{2}{c|}{} \\
\hline
\end{tabular*}

{\footnotesize 
\vspace{1mm}
$^a$ Value 0.56(5) at $N / V$= 8.5 for $\beta$ = 6.4 evolved to $N / V$=3.2
using $\beta$ = 6.0 data. \\  
%[-1ex]
$^b$ Extrapolation sketched in Fig.~\ref{f8}\\
$^c$ Value 0.3 scaled 5.6\% using $a(\sqrt\sigma=$ 440\,MeV).\\
$^d$ From graphs of $N/V$ = 1.81 and 1.43 data. Evolution to 3.2 would reduce
$\rho$  further.\\
$^e$ Value 0.36 scaled 9\% using $a(\sqrt\sigma=$ 440\,MeV). Correlation
function range underestimates average $\rho$ \cite{deforcrand+97}.\\
$^f$ Lattice spacing increased 9\% using $a(\sqrt\sigma=$ 440\,MeV).\\
$^g$ Lattice spacing from hadron masses increased 18\% using $a(\sqrt\sigma=$
440\,MeV).
%***Note leave a blank line to force proper footnote spacing!

}
\end{table*}

\subsection{Vacuum instanton distribution}

Key features and salient results of five studies of instantons in the SU(3)
vacuum are summarized in Table \ref{t1}.

A primary focus has been measuring the distribution of instanton sizes,
$n(\rho)$, and the corresponding average size $\bar\rho$. The fundamental
limitation is the fact that the distribution is modified to some extent by
each cooling procedure. Although in principle, cooling with the Wilson action
eventually allows instantons to shrink and eventually fall through the
lattice, for the small number of cooling steps used with it in these
studies, the effect is inconsequential. A serious problem for all methods, 
however, is the fact that sequential minimization of the action will rotate
the relative color orientation of instanton--anti-instanton pairs into the
attractive direction and bring them together until they annihilate.  Even
over the relatively small range of cooling steps in  Ref.~\cite{smith+98}, for
example, instanton number densities range from 55 fm$^{-4}$ to 3 fm$^{-4}$, and
 Refs.~\cite{chu+94,deforcrand+98} extend down to 0.3 fm$^{-4}$. Continued
evolution would eventually annihilate all the pairs until only instantons or
anti-instantons remain.  The trade-off in all measurements is the uncertainty
in identifying instantons and anti-instantons in the presence of
fluctuations for a small number of steps versus the damage to the original
instanton ensemble caused by a large number of steps.

To compare the various calculations and emphasize
the similarity of their essential features as much as possible,  two
alternative definitions of the average instanton size have been used in
Table \ref{t1}. One definition is the density extrapolated in the number of
cooling steps, $n_c$, to $n_c=0$. This is more representative of the
original, unfiltered vacuum but more subject to pattern recognition errors.
As mentioned previously, it is necessary for APE smearing where instanton
sizes were observed to evolve linearly. Data from 
Refs.~\cite{ivanenko+98,smith+98} are plotted in Fig.~\ref{f8} with linear
extrapolations and the error bars bracketing these extrapolations are listed in
Table
\ref{t1}. The calculation in   Ref.~\cite{chu+94} is effectively an extrapolation,
since the scale calculated from hadron masses evolves nearly linearly with the
number of cooling steps.

\begin{figure}[tp]
\vspace*{-0cm}
$$
\BoxedEPSF{%new-axial-xmgr-bin_bw.eps scaled 400
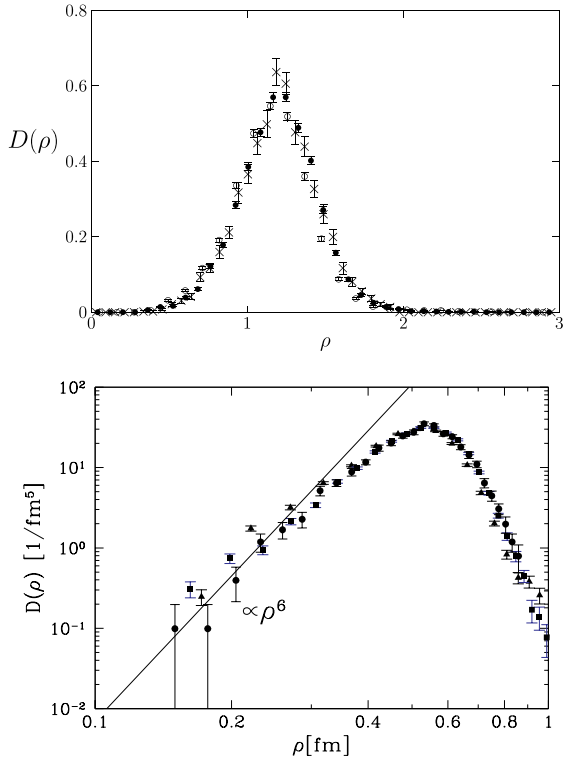 scaled 550}  
$$
\vspace*{-3pc}
\caption{UKQCD instanton size distribution \protect\cite{smith+98}.
In top plot, $\rho=1$ corresponds to 0.45\,fm.
% The upper graph showe distributions at $\beta$ = 6.0, 6.2, and 6.4 for 23,
% 46, and 80 cooling steps denoted by open circles, closed circles, and
% crosses on a scale such that
% $\rho=1$ corresponds to 0.45fm. The lower log-log plot shows apparent
% power law asymptotic behavior.
}
\label{f5} 
\vspace*{-.5cm}  
\end{figure}

The second definition is the continuum limit of comparably filtered
configurations.   Ref.~\cite{smith+98} defines a calibrated number of cooling
sweeps
$n_c(\beta)$ to reach the same density of instantons $N\over V$ at each
$\beta$ and extrapolates linearly in $a^2$ to the continuum limit. This is
consistent with the fact that observables such as the instanton distribution
shown in Fig.~\ref{f5} agree well at $\beta$ = 6.0, 6.2, and 6.4. 

To facilitate comparison in Table \ref{t1}, as indicated in the footnotes,
authors' data were scaled to a lattice spacing determined by the string
tension
$\sqrt\sigma$ = 440 MeV and $\bar\rho_{ N\over V}$ was evaluated as close
to
$N\over V$ = 3.2 fm$^{-4}$ as possible. 
From the table, there is qualitative agreement among all
the calculations and the extrapolated value $\bar\rho_{\rm extrap}$ lies in the
range 0.39 $\pm .05$ fm, consistent with the instanton liquid model, while
the value after cooling to $N\over V$ on the order of 3.2 m$^{-4}$ yields
$\bar\rho_{N\over V}$ substantially larger, in the range 0.54 $\pm .05$ fm.
Although no attempt has been made to calculate $N\over V$ characterizing the
uncooled vacuum, it is clear that the density of instantons identified during
the early stages of cooling is substantially larger than the value 1
fm$^{-4}$ in the instanton liquid model, which should likely be thought of
as an effective density in the size range dominating physical processes. 

\begin{figure}[tbp]
$$
\BoxedEPSF{%new-axial-xmgr-bin_bw.eps scaled 400
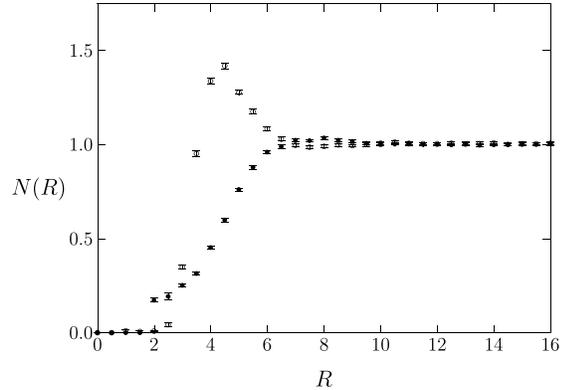 scaled 600} 
$$
\vspace*{-3pc}
\caption{Distribution of like (solid circles) and opposite (open circles)
charges as a function of distance R from a reference charge~\cite{smith+98},
where 1\,fm corresponds to R = 2.2}
\label{f6}  
\vspace*{-.6cm} 
\end{figure}

Each of the studies in Table \ref{t1} also calculated the distribution
of instanton sizes, with similar results. The  UKQCD results~\cite{smith+98},
which had the highest statistics and greatest range of parameters, are shown
in Fig.~\ref{f5}. The upper plot shows the high level of consistency
obtained between comparably cooled configurations with $\beta$ = 6.0, 6.2,
and 6.4 for 23, 46, and 80 cooling steps denoted by open circles, closed
circles, and crosses.
Similar consistency was demonstrated at $\beta$ = 6.0 between
configurations on $16^3\times 48$ and $32^3\times 64$ lattices. 
Full QCD with
${m_{\pi}\over m_{\rho}}$ = 0.82, has a similar
distribution~\cite{ivanenko+98}. The log-log plot~\cite{ringwald+98} of the
UKQCD data shows that the distribution of  instantons grows
roughly as
$\rho^6$ at small $\rho$ as expected in the dilute  approximation,
 falls significantly below $\rho^6$ in the region of
0.3$\,$fm, and decreases at large $\rho$ much more rapidly than 
$\rho^5$ characterizing a frozen coupling constant.

\begin{table}[tp]
\setlength{\tabcolsep}{1pc}
\caption{Topological Susceptibility}
\label{t2}
\begin{tabular*}{\columnwidth}{|l  @{\extracolsep{\fill}} l  l |}
\hline
$\chi^{1/4}$ (MeV) & \quad Method  & Ref.\\
\hline
\multicolumn{3}{|c|}{ } \\[-2ex]
\multicolumn{3}{|c|}{SU(2)}\\[.5ex]
\hline
230 (30) & RG Cycling &   \cite{degrand+97} ${}^{\mathstrut}$\\ 
%DeGrand, A.Hasenfratz, Kov\'acs & 9705009 
220 (6) & APE Smearing & \cite{degrand+98} \\
%DeGrand, Hasenfratz, Kov\'acs & 9711032\\ 
200 (15) & Improved Cooling & \cite{deforcrand+97} \\
%de Forcrand, P\'erez, Stamatescu & 9701012\\ 
198 (8) & APE + Renorm. &  \\
%All\'es, D'Elial, DiGiacomo,  & 9711026\\
            & Geometric + Renorm. &  \cite{alles+97a}\\ 
226 (4) & Spectral Flow &  \cite{edwards98} \\
%Edwards, Heller, Narayan & This meeting\\
\hline
\multicolumn{3}{|c|}{ } \\[-2ex]
\multicolumn{3}{|c|}{SU(3)}\\[.5ex]
%$\chi^{1/4}$ (MeV) & Method & Reference\\
%\multicolumn{3}{|c|}{ } \\
\hline 
187 (14) & Underrelaxed Cooling & \cite{smith+98} ${}^{\mathstrut}$\\ 
%Smith, Tepper &       9801008
192 (5) & APE Smearing &\cite{ahasenfratz+98}\\
% Hasenfratz, Nieter & \\
185 (9) & Improved Cooling & \cite{deforcrand+98} \\
%de Forcrand, P\'erez, Hetrick, &  9802017\\
%             &                             & \quad  Stamatescu & \\ 
175 (5) & APE + Renorm. & \cite{alles+97b} \\
%All\'es, D'Elia, DiGiacomo & {\it NPB\/}    494(1997)2\\ 
197 (4) & Spectral Flow & \cite{edwards98} \\
%Edwards, Heller, Narayanan & This           meeting\\%[1ex] 
\hline
\hline
\multicolumn{3}{|c|}{ } \\[-2ex]
180 & Veneziano-Witten  & \cite{witten79,veneziano79}  \\[.5ex]
\hline
\end{tabular*}
\end{table}

Interesting correlations between instantons have also been
observed~\cite{smith+98}.  Fig.~\ref{f6} shows the distribution of like and unlike
 topological %%% was <toploogical>
 charges as a function of the distance from a reference charge and
clearly reflects the effect of instanton--anti-instanton attraction.
Quantitatively, the average distances to the nearest like and unlike charges
are 0.49 and 0.45 fm respectively, to be compared with the average size 0.56
fm. In a dilute gas, the distribution would be Poisson implying 
$\langle Q^2\rangle = N_I %%% \langle\rangle, not < >
+ N_A$, which only holds for $\rho < 0.5$fm. By all measures, the instanton
distribution is not dilute, and is best thought of as a liquid. In addition,
smaller and larger instantons tend to have opposite charge, with the smaller
instantons having the sign of the total Q, so that the large charges are
overscreened. Many other interesting details are found in~\cite{smith+98}.

\subsection{Topological susceptibility}

It appears that we now 
understand how to define quantities on a  lattice that
correspond to the topological charge in the  continuum limit and
that, as shown in Table \ref{t2}, all methods for calculating the
topological susceptibility are reasonably consistent with each other and with the
Veneziano-Witten formula. 

As emphasized in   Ref.~\cite{alles+97a}, all measurements of the susceptibility
involve both additive and multiplicative renormalization, $a^4\chi^i =Z(\beta)^{-2}
(\chi^i_{\mathop{\rm lattice}}\!\!-\! M(\beta))$. By placing
known smooth configurations on a lattice and heating by Monte Carlo updates, M
was measured for L\"uscher's geometrical charge and both M and Z were
measured for the clover approximation to F\~F with zero and two APE
smearing steps, leading to a consistent value of $\chi$ for all three
calculations~\cite{alles+97a}.  Cooling automatically
removes the additive term and brings Z to unity by removing quantum
fluctuations. It determines the
susceptibility accurately because the total charge is unaffected by pair
annihilation and because, with improved charge operators, the charge
converges to a stable integer after very few cooling steps. Hence, the various
cooling methods yield consistent results as shown. Finally, determination of
the topological charge from spectral flow~\cite{edwards98} also appears to be
consistent with other methods. 

The susceptibility in SU(3) has also been studied using the geometrical
charge~\cite{grandy+97}. The Wilson action yields
$\chi^{1\over 4}$ = 228 MeV, which is expected to be high because of the
additive contribution from dislocations. Calculations with a
renormalization group improved action that suppresses dislocations appear
to have less additive renormalization, but lack of an accurate
measurement of the scale precludes quantitative comparison.

\subsection{Equilibration of topological sectors}

An important  problem in Monte Carlo calculations with dynamical
fermions is ergodically sampling the relevant topological sectors.  Recent hybrid
Monte Carlo calculations~\cite{alles+98} with 3000 to 5000 trajectories on 16$^3
\times$32 lattices using Wilson fermions have demonstrated sufficient
tunneling to achieve equilibration of Q at  $m_{\pi}\over
m_{\rho}$= 0.84, 0.76, and 0.69. At the lightest quark mass, an
approximately Gaussian distribution of Q is observed and the integrated
autocorrelation time for Q, 
$\tau\!= \!$ 54(4), is only moderately larger than  the value $\tau\!= \!$42(4)   for
 other observables.   For $m_{\pi}\over m_{\rho}$ = .56, the mean value of Q did
not equilibrate to 0 within 3,500 trajectories. These results are
consistent with previous HMC calculations with staggered fermions, for which
autocorrelation times of the order of 10 were observed for $m_{\pi}\over
m_{\rho}$ =0.75 and 0.65, but $m_{\pi}\over m_{\rho}$ = 0.57 exhibited little
mobility over 450 trajectories~\cite{alles+96a}. Thus, there is  an
intermediate range of masses down to ${m_{\pi}\over m_{\rho}}\sim$ 0.69 for
which equilibration of topological sectors is achieved for both Wilson and
staggered fermions. 

\section{INSTANTONS AND THE STATIC \\POTENTIAL}

Several puzzles concerning the contribution of instantons to
the static potential have emerged in recent years. A provocative
calculation~\cite{fukushima+98} suggested that an instanton liquid with a
distribution
$\sim
\rho^{-5}$ at large $\rho$ produced a linear confining potential with the
physical string tension. In addition, a confining potential appears to
survive renormalization group cycling for SU(2), but largely disappear when
the instantons identified in the cycled configuration are replaced by
instantons of comparable size and random orientation~\cite{degrand+97}. 

These and related issues are addressed by high statistics
calculations~\cite{brower+98a} of the potentials in an instanton  liquid 
with fixed instanton size and with the size distribution $\rho^6
(\rho_0^{3.5} +
\rho^{3.5})^{-{11\over 3.5}} $, corresponding to a frozen coupling constant . 
The potentials are linear at
 small distance and approach a constant at large distance as expected
 analytically~\cite{callan+78b}. With $\bar\rho \!=\! {1\over 3}$ fm and ${N\over
V}\!=\!
 1 {\rm fm}^{-4}$, both distributions yield the same slope, $\sigma$ = 0.11
GeV/fm, roughly one tenth of the physical
 string tension. Whereas the fixed size potential approaches a constant
already by 1.5 fm, the $\rho^{-5}$ potential is still rising substantially at
3 fm due to the contributions of large instantons.  Because the potential
is roughly proportional to
$N\over V$ and since by scaling, the physical string tension would be
obtained at
$N\over V$ = 1 fm$^{-4}$ by increasing $\bar\rho$ by 10$^{1\over 4}$ to 0.59
fm, one can understand several other results. The peak of the distribution in
 Ref.~\cite{fukushima+98} is at 0.4 fm corresponding to $\bar\rho$ above 0.5
fm, so the slope should be near the physical value. Also, the potential only
extended to 1 fm, so the departure from confinement was not observed.
Similarly, cooled distributions with larger instanton density and $\bar\rho$
are also enhanced.

 The question of what correlations remain in the renormalization group
 cycled configurations yielding either confinement or at least a large slope
 out to 1 fm that are not contained in a comparable random distribution of
 instantons is interesting and should be pursued. One suggestive feature is
 the fact that an instanton liquid as usually constructed does not have Z(N)
 symmetry, whereas confining configurations generated by sampling the
 action automatically have it.  

\section{INSTANTONS AT NONZERO \\TEMPERATURE}

\subsection{Topological susceptibility}

Measurements of the topological susceptibility beyond the critical
temperature confirm the idea that instantons should be suppressed by Debye
screening, where the leading suppression factor~\cite{pisarski+80} is
$e^{-{1\over 3}(2N_c + N_f)(\pi\rho T)^2}$  As at zero temperature, within
one's ability to read numbers from graphs, quenched measurements using one
and two APE smearing steps corrected for additive and multiplicative
renormalization~\cite {alles+97b} agree with analysis of cooled
configurations~\cite{chu+95}.  The cooled measurements showed that the average
instanton size decreased from
$\bar\rho$ = 0.33 fm at low temperature to 0.26 fm at 1.34 T$_C$ and that
with the latter value, Debye screening fit the rapid fall off of the
susceptibility. Comparable cooling calculations with staggered
fermions~\cite{chu+97} showed analogous results, with the instanton size
$\bar\rho$ = 0.44 fm at .75 T$_C$ decreasing to 0.33 fm at 1.3 T$_C$ and
Debye screening with the latter size fitting the susceptibility above T$_C$.
The transition is considerably sharper than in the unquenched case,
consistent with the N$_f$ dependence of the Debye screening.

\subsection{Calorons}

\newcommand {\p}{{\mathcal{P}_{\infty}}}

Gauge fields at nonzero temperature may be  classified by the Polyakov loop
$\mathcal{P}_{\infty}$ = Lim$_{x\to\infty}Pe^{\int_0^{\beta}dtA_0(\vec
x,t)}$, and one usually considers $\p = \pm1$ since fluctuations
around classical solutions with other values produce a non-zero vacuum energy
density~\cite{gross+81}. The periodic instanton, or caloron is given by  
$A_{\mu}(x) = {i\over 2}\bar\eta^3_{\mu\nu}\tau_3
\partial_{\nu}ln \phi$, with $\phi$ a periodic sum of instantons displaced
in the thermal direction~\cite{harrington+78}.  In the limit where the
instanton size
$\rho$ is large compared with $\beta$, the spatial distribution becomes
independent of t, and gauge transformation yields a BPS monopole at the
center of the instanton (and a monopole with opposite charge at
infinity~\cite{rossi79}.

\begin{figure}[tp]
$$
\BoxedEPSF{
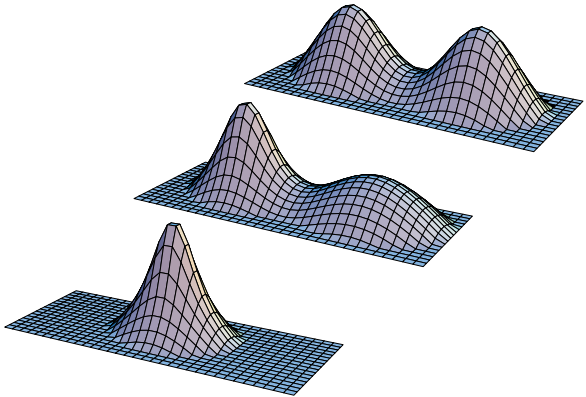 scaled 300}  
$$
\vspace*{-3pc}
\caption{Action density profiles for calorons~\cite{kraan+98} at $\omega = 0$,
${1\over 8}$, and ${1\over 4}$ with $\rho = \beta = 1$.}
\label{f7} 
\vspace*{-.5cm}  
\end{figure}

During the past year, a remarkable set of solutions comprised of
monopoles has been discovered~\cite{kraan+98,lee+98} for nontrivial values of
$\p$ , corresponding to to a periodic array of instantons twisted by
$\p$. The closed form solutions are surprisingly simple. For SU(2)
with ${\p }= e^{2\pi i \omega \tau_3}$ the action density is
tr$F_{\mu\nu}^2 \!= \!\partial^2_{\mu}\partial^2_{\nu}\,$ln$\,\psi$, where
$\psi=$\\$-$cos$(2\pi t) \!+\!c_1c_2 \!+\!{r_1^2+r_2^2+\pi^2\rho^2\over 2 r_1
r_2}s_1s_2\! + \!\pi\rho^2({s_1c_2\over r_1}\!+\!{s_2c_1\over r_2})$, the
positions of two constituent monopoles are denoted $r_m=|\vec x-\vec y_m|$
for m=1, 2,  $|\vec y_2-\vec y_1|=\pi\rho^2$, $\nu_1=2 \omega$, $\nu_2 =
1-2\omega$, $c_m =$ cosh$(2\pi\nu_mr_m)$, $s_m = $ sinh$(2\pi\nu_mr_m)$, and
the solution is scaled to $\beta=1$. General expressions are given for
tr$F_{\mu\nu}^2$ and
$A_{\mu}(x)$ for SU(N)  in   Ref.~\cite{kraan+98}. 

In the SU(2) case, for $\rho$ large relative to $\beta$ the the solution
separates into two lumps separated by $\pi\rho^2\over \beta$ with monopole
charges $\pm 1$, masses
$8\pi^2\nu_m\over \beta$, and  widths $\sim {1\over \nu_m}$. In the limit
of large $\rho \over \beta$ the lumps become well separated, spherically
symmetric, and time independent.  Since they are  self-dual, they
are therefore BPS monopoles. As $\omega$
approaches 0 or $1\over2$, the second lump becomes negligible and the
solution becomes the spherically symmetric Harrington-Shepard caloron. 
The action densities plotted on logarithmic
scales in Fig~\ref{f7} at t=0 show the separation into lumps at $\rho =
\beta$. For small $\rho\over \beta$, the caloron approaches a single
instanton solution. In the maximal abelian gauge, as $\rho$ increases from
0 to $\infty$, the small monopole loops associated with instantons grow.
When these loops extend from 0 to $\beta$, they merge
with their periodic extensions to produce pairs of monopole lines which
eventually become the well-separated BPS monopoles~\cite{brower+98b}. In the
case of SU(N), the solution generalizes to N monopole lumps, and results are
shown in   Ref.~\cite{kraan+98}.

These novel finite temperature solutions provide interesting examples of
gauge fields with topological charge which are built out of monopoles
without gauge fixing, and we may hope that they will lead us a step closer to
analytical insight into the role of monopoles in QCD.

\section{SUMMARY AND OPEN PROBLEMS}

We have come a long way in exploring  instanton physics on the lattice.  
The behavior of two-point functions, the similarity of cooled and
uncooled calculations, and the contributions of zero modes provide
substantial evidence that instantons play a major role in the physics
of light hadrons. There has been significant progress in calculating quark
zero modes, observing localization, extracting the instanton content of the
vacuum, calculating the topological susceptibility, and discovering analytic
solutions with monopole constituents of instantons. Salient problems for
further investigation include reducing the ambiguity in the instanton
distribution in the vacuum, removing the effect of doublers in spectral flow
analyses, understanding the role of instantons in confinement, and
especially studying full QCD with light chiral fermions, where the
contributions of instantons and zero modes become most significant.

\subsection*{Acknowledgments}

It is a pleasure to acknowledge Richard
Brower, Dong Chen, Ming Chu, Philippe de Forcrand, Tom DeGrand, Anna
Hasenfratz, Suzhou Huang, Taras Ivanenko, Andrew Pochinsky, Edward Shuryak,
and Uwe-Jens Wiese for enlightening discussions and insights.

\end{document}